\documentclass{article}
\usepackage{setspace}

\usepackage{graphicx}% Include figure files
\usepackage{dcolumn}% Align table columns on decimal point
\usepackage{bm}% bold math
\usepackage{amssymb}
\usepackage{amsmath}

\begin{document}

%\preprint{APS/123-QED}

\title{Classical Universe emerging from quantum cosmology without horizon and flatness problems}
\author{M. Fathi$^1$,\,\, S. Jalalzadeh$^1$\thanks{s-jalalzadeh@sbu.ac.ir}\,\,\,and\,\,\,P.V.
Moniz$^{2,3}$\thanks{pmoniz@ubi.pt}\\
$^1${\small Department of Physics, Shahid Beheshti University, G. C., Evin, Tehran, Iran}\\
$^2${\small Centro de Matem\'{a}tica e Aplica\c{c}\~{o}es- UBI, Covilh\~{a}, Portugal}\\
$^3${\small Departmento de F\'{\i}sica, Universidade da Beira Interior, 6200 Covilh\~{a}, Portugal}}

\maketitle
\begin{abstract}
We apply the complex de Broglie-Bohm formulation of quantum mechanics \cite{Complex} to  a spatially closed 
 homogeneous and  isotropic  early Universe whose matter content are radiation
and dust perfect fluids. We then show  that  an expanding  classical Universe can emerge from an oscillating (with complex scale factor) quantum Universe without  singularity. Furthermore, the Universe obtained in this process has no horizon or flatness problems.
\end{abstract}

%\pacs{04.20.Cv, 04.60.-m, 98.80.Bp, 98.80.Qc}% PACS, the Physics and Astronomy

\section{Introduction}

In canonical quantum cosmology, the wave function of Universe is obtained
from the Wheeler-DeWitt (WDW) equation which is time independent and consequently we have no quantum dynamics.
Quantum mechanically speaking, as we know, the Copenhagen interpretation
applied to cosmology  has some serious  problems: The impossibility of a clear division of the total Universe
into the observer (who measures) and the observed makes difficult to interpret the wave function of Universe. Moreover, assuming the existence of only one observable Universe, the interpretation
of the absolute square of the wave function as a probability density is impossible.
 To find a solution to  above mentioned problems via quantum
cosmology, the straight and direct way could be the  de Broglie-Bohm (dBB), or causal stochastic, interpretation of quantum cosmology. 
 The dBB interpretation is favorable, especially for a quantum theory of
cosmology, because this interpretation is able to resolve the above mentioned conceptual problems of quantum cosmology \cite{Bohmian}.
However, we have a problem  in using  dBB interpretation in quantum
cosmology. It cannot describe the  trajectories and non-zero velocities for real wave functions in the minisuperspace (see next section for
more details).

In this paper, we propose to look at the problems of standard cosmology  from a different and novel quantum cosmological perspective. In recent years, the complex de Broglie-Bohm (CdBB) formulation of quantum
mechanics has been developed
as a new alternative interpretation of quantum mechanics \cite{Complex}.
It is based on the quantum Hamilton-Jacobi formalism introduced 
by Leacock and Padgett \cite{Leacock}.  One of the advantages of this model is that it does not face the problem of stationarity of particles in bound states, encountered in the  dBB representation
\cite{John}.
The CdBB formulation can be introduced as follows. We employ $\Psi=e^{iS(q^\mu)}$,
$S\in\mathbb{C}$, in the corresponding wave equation of the quantum system to obtain a single CQHJ equation. Since the action  $S$ is complex valued and time
remains real valued, the position and conjugate momentum of particles are
complex valued. In this description \cite{Complex}, the transition from a quantum regime
to the corresponding classical world occurs for simultaneous  very large values of position and quantum numbers of system \cite{Har}, where the quantum force  disappears and the particle's motion is entirely governed by the classical equation of motion.

In this paper we will investigate, in the CdBB framework, the quantum cosmology of a simple
closed FLRW Universe, filled with   radiation and dust fluids. In section II, we develop the CQHJ interpretation of our model. We obtain the  state dependent quantum cosmological solutions with complex trajectories in complex minisuperspace in section III. In section IV, we show that for large values of the scale factor and state number $n$, the model emerges into a classical cosmology, without the horizon and flatness problems.

\section{Complex Bohmian quantum cosmology in minisuperspace}

Simple   cosmological models are achieved by considering a class of models in which all but finite number
of degrees of freedom of metric and matter fields are ``frozen''. This is
most commonly achieved by restricting the fields to be homogeneous, so that
the line element of spacetime is given by
\begin{eqnarray}\label{ss1}
ds^2=-N^2(t)dt^2+h_{ij}(t,x^j)dx^idx^j,\hspace{.3cm}i,j=1,2,3.
\end{eqnarray}
where $N(t)$ is the lapse function and the 3-metric $h_{ij}$ are restricted to be homogenous. Using the above line
element and also assuming the homogeneity of matter fields, the Lagrangian of Einstein-Hilbert plus the matter fields reduce to the minisuperspace form \cite{Hall}
%\begin{widetext}
\begin{eqnarray}\label{ss2}
%\begin{array}{cc}
\mathcal L=\frac{1}{2N}f_{\alpha\beta}(q^\mu)\dot q^\alpha\dot q^\beta-NU(q^\mu),\,\,\,\,\,\alpha,\beta=0,1,2,...,n-1,
%\end{array}
\end{eqnarray}
%\end{widetext}
where $f_{\alpha\beta}$ is the metric of minisuperspace (a reduced version
of the full DeWitt metric) with indefinite signature $(-,+,+,+,...)$, $q^\alpha(t)$
denotes local finite coordinates of minisuperspace and $U(q^\mu)$ is a particularization
of $-\sqrt hR^{(3)}(h_{ij})+V(Matter)$, where $V(Matter)$ represents the
potential terms coming from matter degrees of freedom. Note that sometimes
it is convenient to scale the lapse function in terms of other  minisuperspace
metric elements (see next section). To obtain the canonical Hamiltonian, we first
define canonical momentum
\begin{eqnarray}\label{ss3}
p_\alpha=\frac{\partial\mathcal L}{\partial\dot q^\alpha}=f_{\alpha\beta}\frac{\dot
q^\beta}{N}.
\end{eqnarray}
Hence, the canonical Hamiltonian is given by
\begin{eqnarray}\label{ss4}
H_c=p_\alpha\dot q^\alpha-\mathcal L=N\left[\frac{1}{2}f^{\alpha\beta}p_\alpha
p_\beta+U(q^\mu)\right]:=N\mathcal H,
\end{eqnarray}
where $f^{\alpha\beta}$ is the inverse metric on minisuperspace. The Hamilton
equations
\begin{eqnarray}\label{ss5}
\begin{array}{cc}
\dot q^\alpha=\frac{\partial H_c}{\partial p_\alpha}=Nf^{\alpha\beta}p_\beta,\\
\dot p_\alpha=-\frac{\partial H_c}{\partial q^\alpha}=-N\left(\frac{1}{2}f^{\mu\nu}_{,\alpha}p_\mu
p_\nu+U_{,\alpha}\right),
\end{array}
\end{eqnarray}
leads us to the field equations
\begin{eqnarray}\label{ss6}
\frac{1}{N}\frac{d}{dt}\left(\frac{\dot q^\alpha}{N}\right)+\frac{1}{N^2}\Gamma^\alpha_{\mu\nu}\dot
q^\mu\dot q^\nu+f^{\alpha\beta}U_{,\beta}=0,
\end{eqnarray}
where $\Gamma^\alpha_{\mu\nu}$ are the components of a Christoffel connection
compatible with metric  $f$. In addition, the gauge freedom on choosing lapse function
leads to the following weak equation for super-Hamiltonian
\begin{eqnarray}\label{ss7}
\mathcal H=\frac{1}{2}f^{\alpha\beta}p_\alpha p_\beta+U(q^\mu)\approx0.
\end{eqnarray}

\subsection{Canonical quantization}
The canonical quantization of this model is accomplished in the coordinate representation,
$q^\alpha=q^\alpha$, $p_\alpha=-i\partial_\alpha$ and demanding that the
time independent wave function $\Psi(q^\mu)$ is annihilated by the self-adjoint
operator corresponding to the Hamiltonian constraint (\ref{ss7}), which gives
the WDW
equation
\begin{eqnarray}\label{ss8}
\mathcal H\left(q^\alpha,-i\partial_\alpha\right)\Psi(q^\mu)=0.
\end{eqnarray}
To solve the operator ordering problem, we should assume that the minisuperspace
metric
part of WDW equation is covariant under general coordinate transformations in minisuperspace
and is also  conformal invariant \cite{SS1}. Consequently,
the WDW equation will be
\begin{eqnarray}\label{ss9}
\left[-\frac{1}{2}\square+\xi\mathcal{R}+U(q^\mu)\right]\Psi(q^\mu)=0,
\end{eqnarray}
where $\mathcal R$ is the Ricci scalar associated to minisuperspace Semi-Riemannian
manifold $(f,\nabla)$, $\xi=-\frac{n-2}{8(n-1)}$ for $n\geqslant2$ and $\square=f^{\alpha\beta}\nabla_\alpha\nabla_\beta=\frac{1}{\sqrt{-h}}\partial_\alpha(\sqrt{-h}f^{\alpha\beta}\partial_\beta)$
is the D'Alembert operator. Moreover, the covariantly conserved $n$-current corresponding
to the WDW equation is given by
\begin{eqnarray}\label{ss10}
J_\mu=\frac{1}{2i}\left(\Psi^*\nabla_\mu\Psi-\Psi\nabla_\mu\Psi^*\right).
\end{eqnarray}
Note that the WDW equation is a Klein-Gordon type and consequently the probability
measure constructed from the above current suffers the same difficulties
with negative probabilities in the usual Klein-Gordon equation.

\subsection{de Broglie-Bohm quantum cosmology}
Before we proceed further, some comparisons with dBB approach
to quantum cosmology \cite{SS2} will be helpful  to explain the necessity of extending
the concept of quantum trajectory to complex domain. 

The WDW equation (\ref{ss9}) is separable by means of the general complex
assumption (de Broglie ansatz)
\begin{eqnarray}\label{ss11}
\Psi(q^\mu)=R_B(q^\mu)e^{iS_B(q^\mu)},\hspace{.3cm}R_B\,\, \text{and}\,\, S_B\in\mathbb{R}.
\end{eqnarray}
The subscript ``B'' is introduced to highlight the obtained results from
dBB  with CQHJ approach. Substituting (\ref{ss11}) into WDW equation (\ref{ss9}) and separating into
real and imaginary parts, gives two coupled non-linear partial differential
equations respectively
\begin{eqnarray}\label{ss12}
\frac{1}{2}f^{\alpha\beta}\nabla_\alpha S_B\nabla_\beta S_B+\xi\mathcal R(q^\mu)+U(q^\mu)+Q_B(q^\mu)=0,
\end{eqnarray}
\begin{eqnarray}\label{ss13}
\nabla_\alpha J^\alpha=\nabla_\alpha(f^{\alpha\beta}R^2_B\nabla_\beta S_B)=0,
\end{eqnarray}
where 
\begin{eqnarray}\label{ss14}
Q_B(q^\mu):=-\frac{\square R_B}{2R_B}=-\frac{1}{2}\frac{\square|\Psi|}{|\Psi|},
\end{eqnarray}
is the quantum potential. The assumption introduced by the dBB approach is
that we have well-defined location $q^\alpha$ together with $n$-momentum
\begin{eqnarray}\label{ss15}
p_\alpha:=\nabla_\alpha S_B=\frac{f_{\alpha\beta}}{N}\dot q^\beta.
\end{eqnarray}
The lapse function is introduced in the definition of momentum because of gauge reparameterization freedom of general relativity. It is obvious that
for real wave functions $S=0$ and consequently the $n$-momentum (\ref{ss15})
vanish.

\subsection{Complex quantum Hamilton-Jacobi cosmology}
The  CQHJ or CdBB  mechanics is one of the nine formulations \cite{St} of 
quantum mechanics, developed along the lines of the classical Hamilton-Jacobi theory. Indeed, it not only provides
an alternative interpretation of  quantum mechanics but may also serve as a powerful tool to solve quantum mechanical problems \cite{To}.
The starting point of  the CQHJ formalism of quantum mechanics,
instead of (\ref{ss11}), is using the following ansatz \cite{Leacock}
\begin{eqnarray}\label{ss16}
\Psi(q^\alpha)=e^{iS(q^\alpha)},\hspace{.3cm}q^\alpha \,\,\text{and}\,\,S(q^\alpha)\in \mathbb{C},
\end{eqnarray}
where the wave function and the phase are analytically
extended to the complex plane by replacing real coordinates $q^\alpha$ with
 complex coordinates, $q^\alpha=q^\alpha_R+iq^\alpha_I$ { though its value will be (physically) meaningful only along the real axis \cite{Book, book2},}
and keeping time (and lapse function)  real valued.  Substituting this new
ansatz in the WDW equation (\ref{ss9}) yields a single equation, known as CQHJ equation
\begin{eqnarray}\label{ss17}
\frac{1}{2}f^{\alpha\beta}\nabla_\alpha S\nabla_\beta S+\xi{\mathcal R}(q^\mu)+U(q^\mu)+Q(q^\mu)=0,
\end{eqnarray}
where the new complex quantum potential is given by
\begin{eqnarray}\label{ss18}
Q:=\frac{1}{2i}\square S=-\frac{1}{2}\left(\frac{\square\Psi}{\Psi}-\frac{f^{\alpha\beta}\nabla_\alpha\Psi\nabla_\beta\Psi}{\Psi^2}\right),
\end{eqnarray}
brings all quantum effects into the CQHJ
formalism. However, this quantity is not the same as the
Bohm quantum potential, defined in (\ref{ss14}).
Note that there is no expansion in powers of $\hbar$ in the derivation and
Eq.(\ref{ss17}) is exact. In analogy to standard Bohmian mechanics, complex quantum trajectories can
also be defined by analytic continuation of (\ref{ss15}) to the complex plane
as \begin{eqnarray}\label{ss19}
p_\alpha:=\nabla_\alpha S(q^\mu)=\frac{1}{N}f_{\alpha\beta}\dot q^\beta,\hspace{.3cm}p_\alpha\in\mathbb{C},
\end{eqnarray}
{ Therefore, the main novelty of the CdBB formulation is that now the
guidance equation is related to a new complex action function,
$S$, and not only to the real part of wave function.
The relationship between the Bohmian momenta (\ref{ss15}) and its complex counterpart is
\begin{eqnarray}\label{mom}
p_\alpha=p_\alpha^{(B)}-\frac{i}{R_B}\nabla_\alpha R_B,
\end{eqnarray}
 This expression explains why it is possible to observe non-vanishing  momenta in cases where the Bohmian momenta, $p_\alpha^{(B)}$, vanishes. In fact, the Bohmian trajectories defined in Eq.(\ref{ss15}) only carry information about the dynamics of quantum flow. But, complex quantum trajectories defined in Eq.(\ref{mom}) also include information about the probability, because of following relation between the complex and Bohmian action functions
\begin{eqnarray}\label{actions}
S=S_B-i\ln R_B.
\end{eqnarray} 
Therefore, the complex dynamics explains how to get the correct momentum distribution.}

Eq.(\ref{ss19}) is invariant under time reparametrization. To obtain the corresponding field equations, we differentiate $n$-momentum
defined above with respect to cosmic time $t$, which gives $\frac{dp_\mu}{dt}=\dot
q^\alpha\partial_\alpha\partial\mu S=\dot
q^\alpha\partial_\mu\nabla_\alpha S$. Now, differentiation with respect to
cosmic time
of Eq.(\ref{ss17}) and using the
second
equality in Eq.(\ref{ss19}), we obtain
%\begin{widetext}
\begin{eqnarray}\label{ss20}
%\begin{array}{cc}
\frac{1}{N}\frac{d}{dt}\left(\frac{\dot q^\alpha}{N}\right)+\frac{1}{N^2}\Gamma^\alpha_{\mu\nu}\dot
q^\mu\dot q^\nu+f^{\alpha\beta}(U+Q+\xi\mathcal R)_{,\beta}=0,
%\end{array}
\end{eqnarray}
%\end{widetext}
which is the extension of classical field equations (\ref{ss6}) to the complex
quantum minisuperspace. Furthermore, Eqs.(\ref{ss17}) and (\ref{ss19}) give us the complex quantum  super-Hamiltonian
constraint
%\begin{widetext}
\begin{eqnarray}\label{ss21}
%\begin{array}{cc}
\mathcal H=\frac{1}{2}f^{\alpha\beta}p_\alpha p_\beta+\frac{1}{2i}f^{\alpha\beta}\nabla_\alpha p_\beta+\xi\mathcal
R(q^\mu)+U(q^\mu)=0,
%\end{array}
\end{eqnarray}
%\end{widetext}
which is a Riccati-like PDE.

The wave function (\ref{ss16}) is invariant with respect
to a change of its phase $S(q^\mu)$ by an integer multiple of $2\pi
$. Consequently, the
definition of momentum (\ref{ss19}) gives
\begin{eqnarray}\label{aa1}
\oint_C p_\mu dq^\mu=nh=2\pi n, \hspace{.3cm}n=1,2,3,...\,\,,
\end{eqnarray}
as a condition of compatibility between the CQHJ equation (\ref{ss21}) and WDW equation (\ref{ss9}). Here, $C$ is a counter clockwise contour in the complex configuration space, enclosing the real
line between the classical turning points.
 Unlike the real-valued Eq.(\ref{ss12}), the CQHJ equation (\ref{ss17}) contains all of the information present in  wave function of Universe (\ref{ss16}). Some authors have claimed that the CQHJ formalism is  more fundamental
than the dBB interpretation  \cite{SS3}. Moreover, there does not exist an obvious
probability flux continuity equation in the CQHJ formalism as opposed to the coupled equations for real phase and real amplitude in conventional
dBB interpretation. However, the most significant difference
arises from the fact that for bound states and the real wave functions, the predictions from  Bohmian
mechanics acquire a new context:  For wave functions whose
space part is real, the action function in dBB interpretation, $S_B$, is constant and consequently the velocity field is zero everywhere. { In fact, in Bohmian interpretation
of ordinary quantum mechanics, for a stationary bound state, since the $R_B$-amplitude
defined in (\ref{ss11}) is time independent, the  continuity implies that
the Bohmian phase, $S_B$, is constant. To solve this problem Floyd \cite{Floyd} considered
that for this kind of states, the Bohmian phase could be separated into space
and time parts, $S(x,t)=W(x)-Et$, where $W(x)$ is the reduced Bohmian action
function. Then, Floyd defined the energy dependent modified potential by
$U=V+Q_B$, where $V$ is the ordinary original potential of wave equation. He pointed out that for a given eigenvalue of Schr\"odinger equation, there
is an infinite number of modified potentials $U_1,\,U_2\,...$, and associated
with each of these potentials is a trajectory $x_1,\,x_2\,...\,$. But then
the Floydian microstates, $\{U_i,x_i\}$, do not arise directly from Schr\"odinger
equation and this description is not equivalent to the original wave equation,
regarding this fact that the microstates provide new dynamical information
that is not contained in Schr\"odinger
equation \cite{Book}. But as we know, in quantum gravity and also in quantum cosmology, the general
covariance indicate that the wave function is time independent (time problem).
Consequently, it is clear that  this resolution of problem is not working in quantum
cosmology.}  This is  an undesirable feature in dBB interpretation, which claims to make the theory perceivable and
causal \cite{John}. On the other hand, in CQHJ formulation we can obtain
a general velocity field.

 Let us further elaborate on the difference of these two approaches with a very simple example
from non-relativistic quantum mechanics. Consider the particle in a box model (the infinite square well) which describes a particle free to move in a small one-dimensional space, $0\leqslant q\leqslant L$, surrounded by impenetrable barriers which is a simple
 model mainly used  to illustrate the differences between classical and quantum mechanics. The space part of the wave function is
given by $\psi_n(q)=\sqrt{\frac{2}{L}}\sin(\frac{n\pi q}{L})$, where $n$ is
an integer quantum number. In Bohmian mechanics, comparing this wave function with (\ref{ss11}) gives us $S_B=0$ and therefore the momentum of a particle
defined by (\ref{ss15}) will be zero. On the other hand, in CQHJ formulation, according to the definition (\ref{ss16}) the action function is given by $S=\ln(\sin(\frac{n\pi q}{L}))$. Therefore, using
(\ref{ss19}) the momentum of particle will be $p=\frac{dS}{dq}=m\frac{dq}{dt}=-\frac{in\pi\hbar}{L}\cot(\frac{n\pi
q}{L})$, where $m$ is the mass of particle. By analytic continuation  into complex space, $q=q_\text{R}+iq_\text{I}$, where $q_\text{R},q_\text{I}\in\mathbb{R}$ and solving the above
differential equation, we obtain
\begin{eqnarray}\label{ss22}
\begin{cases}
\cos(\frac{n\pi q_\text{R}}{L})\cosh(\frac{n\pi q_I}{L})=C\cos(\frac{n^2\pi^2\hbar
t}{mL^2}),\\
\sin(\frac{n\pi q_\text{R}}{L})\sinh(\frac{n\pi q_I}{L})=C\sin(\frac{n^2\pi^2\hbar
t}{mL^2}),
\end{cases}
\end{eqnarray}
where $C$ is a constant of integration. Hence, the particle have a well
defined complex quantum trajectory with real and imaginary parts satisfying
the above equations. Furthermore, the
quantum Hamiltonian of particle will be
%\begin{widetext}
\begin{eqnarray}\label{ss23}
%\begin{array}{cc}
H=\frac{1}{2m}(\frac{dS}{dq})^2+Q=\frac{p^2}{2m}+\frac{n^2\pi^2\hbar^2}{2mL^2\sin^2(\frac{n\pi
q}{L})}=\frac{n^2\pi^2\hbar^2}{2mL^2},
%\end{array}
\end{eqnarray}
%\end{widetext}
where the last equality is obtained by replacing the corresponding complex momentum, $p=-\frac{in\pi\hbar}{L}\cot(\frac{n\pi
q}{L})$ .
Let us now examine the classical limit. For very large values of the quantum
number $n$,  using the approximate relations $\cosh(\frac{n\pi |q_I|}{L})=\sinh(\frac{n\pi |q_I|}{L})\simeq\frac{1}{2}\exp(\frac{n\pi |q_I|}{L})$ for $n\rightarrow\infty$, the explicit solution of the coupled equations (\ref{ss22}) will
be
\begin{eqnarray}\label{ss24}
q_\text{I}=\frac{L}{n\pi}\ln(2C)\simeq0,\hspace{.3cm}q_\text{R}=-\frac{n\pi\hbar}{mL}t=p_ct,
\end{eqnarray}
where $p_c$ is the classical  momentum of particle. Therefore, for large
values of quantum number $n$, the imaginary part of trajectory and momentum
will be disappear and
we will have a classical particle with real trajectory.
The transition from quantum mechanics (CQHJ) to classical 
mechanics occurs when the motion of the particle falls entirely on the real subspace.

The wave function of Universe is real valued in many minisuperspace models of Universe \cite{Fa}. To obtain a Bohmian interpretation for these models, the
usual procedure  is constructing a wave packet by superposition of eigenstates 
\cite{Fa}. But it is not clear that the hidden symmetries of a model gives us a permission in general
to construct such wave packets \cite{Man}. On the other hand, the CQHJ interpretation
gives us latitude to obtain causal interpretation even for 
real wave functions of Universe.

\section{FLRW cosmology with a perfect fluid (dust and radiation)}

Let us  consider a closed homogenous and isotropic Universe with line
element
  \begin{eqnarray} \label{line}
 \begin{array}{c}
 ds^2= -N^2 (\eta^*)d{\eta^*} ^2 +a^2 (\eta^*) d{\Omega^2}_{(3)}, 
 \end{array}
 \end{eqnarray}
 where $ N(\eta^*)$ denotes the lapse function, $a(\eta^*)$ is the scale factor  and $d{\Omega^2}_{(3)}$ is the standard line element of unit 3-sphere. The action functional that consists of a gravitational part
and a matter part when the matter field is considered as a
perfect fluid is given by
 \begin{eqnarray} \label{functional}
 \begin{array}{ccc}
\mathcal A=\frac{1}{2} {M^2_\text{Pl}} \int{\sqrt{-g} R\ d^4x}\\
 \\+{M^2_\text{Pl}}\int_{\partial{\mathcal M}} {\sqrt{g^{(3)}} K d^3 x} -\int_{\mathcal M} {\sqrt{-g} \rho  d^4 x},\\
 \end{array}
 \end{eqnarray}
 where $M_\text{Pl}^2=\frac{1}{{8\pi G}}$ is the reduced Planck mass in natural units, $K$ is the trace of the extrinsic curvature of the spacetime boundary, $\rho$ is the total density of matter content of universe and $\mathcal M$ represents the manifold of the spacetime with boundary $\partial{\mathcal M}$ .
Let us also 
 define some useful quantities. 
If we assume a Universe filled with mixture of noninteracting dust, $\rho_m$, and radiation, $\rho_\gamma$, then the total energy density will be
\begin{eqnarray}\label{rr11}
\rho=\rho_m+\rho_\gamma=\rho_{mi}\left(\frac{a}{a_i}\right)^{-3}+\rho_{\gamma
i}\left(\frac{a}{a_i}\right)^{-4},
\end{eqnarray}
where $\rho_{mi}$ and $\rho_{\gamma i}$ denote the energy density
of dust and radiation respectively, at initial time $t_i$ when the scale factor is $a_i=a(t_i)$.
  Setting the initial time as GUT time, $t_i=t_\text{GUT}$, the total energy
density (\ref{rr11}) can be rewrite as
\begin{eqnarray}\label{density}
 \rho=3H_{g}^2M_\text{Pl}^2 \left(\Omega_{m}\left(\frac{a}{a_{g}}\right)^{-3}+\Omega_{\gamma}\left(\frac{a}{a_{g}}\right)^{-4}\right),
 \end{eqnarray}
  where $H_{g}$ and $a_{g}$ are the Hubble parameter and scale factor
  of the Universe at the GUT comoving time, $t_g$. We also define the density parameter
  of dust and radiation at GUT epoch as $\Omega_m=\Omega_m(t_g)=\rho_m(t_g)/(3H^2_gM^2_\text{Pl})$ and $\Omega_\gamma=\Omega_\gamma(t_g)=\rho_m(t_g)/(3H^2_gM^2_\text{Pl})$. In addition,  if we redefine the scale factor, the lapse function and time coordinate as
\begin{eqnarray}\label{scale}
\begin{cases}
q=\frac{a}{a_{g}}-\frac{\Omega_m}{2|\Omega_k|},\\
N=\frac{a}{a_{g}}\tilde N,\\
d\eta=H_{g}d\eta^*,
\end{cases}
\end{eqnarray}
and introduce the following parameters
\begin{eqnarray}\label{pp}
\begin{cases}
M=\frac{12\pi^2M_\text{Pl}^2}{H_{g}^2|\Omega_k|^\frac{3}{2}},\,\,\,\omega=\sqrt{|\Omega_k|},\\

\mathcal E=\frac{M}{2}\left[\frac{\Omega_m^2}{4|\Omega_k|}+\Omega_\gamma\right],
\end{cases}
\end{eqnarray}
where $\Omega_{k}=-\frac{1}{a_{g}^2H_{g}^2}$ denotes spatial curvature density, at the GUT epoch,  then the total Lagrangian of the model in one-dimensional minisuperspace
will be
\begin{eqnarray}\label{third lag}
\mathcal L=\frac{1}{2}\frac{M}{\tilde N}\dot q^2-\frac{1}{2}M\tilde N\omega^2q^2+\tilde
N\mathcal E,
 \end{eqnarray}
 { where a dot denotes derivative respect to the $\eta$. The conjugate momentums
 of  $q$ and the lapse function $\tilde N$ are
 \begin{eqnarray}\label{mo}
 p=\frac{M}{\tilde N}\dot q,\,\,\,\,\,\,p_{\tilde N}=0.
 \end{eqnarray} 
 The canonical Hamiltonian (\ref{ss4}) for this model will be
 \begin{eqnarray}\label{H}
 H_c=\dot qp+p_{\tilde N}\dot{\tilde N}-\mathcal L=\tilde N\left(\frac{p^2}{2M}+\frac{1}{2}M\omega^2 q^2 -\mathcal E\right).
 \end{eqnarray}
 Because of the existence of constraint $p_{\tilde N}=0$, the Lagrangian
 is singular. Hence, the total Hamiltonian could be constructed by adding
 to the Hamiltonian (\ref{H}) the primary 
  constraint, multiplicated by an arbitrary function of time, $\lambda(\eta)$
\begin{eqnarray}\label{ht}
H_T=N\left(\frac{p^2}{2M}+\frac{1}{2}M\omega^2 q^2 -\mathcal E\right)+\lambda
p_{\tilde N}.
\end{eqnarray}  
The requirement that the primary constraint, $p_{\tilde N}=0$, must hold
during the evolution means that
\begin{eqnarray}\label{r}
\dot p_{\tilde N}=\{p_{\tilde N}, H_T\}\approx0.
\end{eqnarray} 
 Eqs.(\ref{ht}) and (\ref{r}) lead us to the secondary constraint
\begin{eqnarray} \label{hamiltoni}
\mathcal{H}=\frac{p^2}{2M}+\frac{1}{2}M\omega^2 q^2 -\mathcal E\approx0,
\end{eqnarray}
which is the weak equation for super-Hamiltonian defined in Eq.(\ref{ss7}).}
\subsection{Brief discussion of the classical minisuperspace}
 The Hamilton equations of motion (\ref{ss5}), in the gauge $\tilde N=1$ will be
\begin{eqnarray}\label{cl1}
\dot q=\frac{p}{M},\,\,\,\,\,\,\,
\dot p=-M\omega^2q,
\end{eqnarray}
which lead us to solution $q=A\cos(\omega\eta+\theta)$, where $A$ and $\theta$ are
constants of integration.  The super-Hamiltonian constraint (\ref{hamiltoni}) fixes the value
of $A$ as $A=\frac{1}{\omega}\sqrt{\frac{2\mathcal E}{M}}$. 
If we assume
that the initial singularity occurs at $\eta=0$, and by using the relation  $a_gH_gdt=ad\eta$
defined in (\ref{scale}) between cosmic time $t$ and conformal time $\eta$, the scale factor in terms
of comoving cosmic time $t$ will be
\begin{eqnarray}\label{cl2}
\begin{cases}
a(t)=\frac{a_\text{Max}}{1-\sec(\theta)}\left[\sec(\theta)\cos(\omega\eta+\theta)-1\right],\\
t=\frac{\Omega_m}{2H_g|\Omega_k|}\left[\eta-\frac{1}{\omega}\sec(\theta)\sin(\omega\eta+\theta)\right],
\end{cases}
\end{eqnarray}
where $\cos(\theta):=-\frac{\Omega_m}{2\sqrt{|\Omega_k|}}\sqrt{\frac{M}{2\mathcal
E}}$ and $a_\text{Max}:=\frac{a_g}{\sqrt{|\Omega_k|}}\sqrt{\frac{2\mathcal E}{M}}+\frac{a_g\Omega_m}{2|\Omega_k|}$ is the maximum value of scale factor.
It
is easy to find that at the GUT epoch, the super-Hamiltonian constraint (Friedmann
equation) reduce
to the well known relation between energy density parameters\begin{eqnarray}\label{yy}
\Omega_\gamma+\Omega_m+\Omega_k=1.
\end{eqnarray}  
    
\subsection{FLRW Quantum cosmology with a perfect fluid}

The standard canonical quantization of this simple model is accomplished
straightforwardly in the coordinate representation $q=q$ and $p=-i\frac{d}{d
q}$. Then the Hamiltonian constraint (\ref{hamiltoni})
becomes the WDW equation for the wave function of the
Universe,
\begin{eqnarray} \label{wdw}
\begin{array}{c}
\Big{(}-\frac{1}{2M}\frac{d^2}{dq^2}+\frac{1}{2}M\omega^2 q^2\Big{)}\Psi_n(q)=\mathcal
E_n\Psi_n(q).
\end{array}
\end{eqnarray}
The eigenvalues and normalised eigenfunctions are
\begin{eqnarray} \label{eigen}
\begin{cases}
\mathcal E_n={(}n+\frac{1}{2}{)}\omega,\\
\psi_{n}{(q)}=\frac{1}{\sqrt{2^n n!}}\Big{(}\frac{M\omega}{2}\Big{)}^\frac{1}{4}e^{-\frac{M\omega}{2}q^2}H_{n}{(}\sqrt{M\omega}q{)},
\end{cases}
\end{eqnarray}
where $H_{n}(x)$ denotes the Hermite polynomials. Substituting ${\mathcal
E}$ and $\omega$ defined by (\ref{pp}) into the eigenvalue equation obtained
in Eq.(\ref{eigen}), we obtain the following relation between energy density parameters 
\begin{eqnarray}\label{mama}
\frac{1}{|\Omega_k|^2}\left(\frac{\Omega_m^2}{4|\Omega_k|}+\Omega_\gamma\right)=\frac{H_g^2}{6\pi^2M_\text{Pl}^2}(n+\frac{1}{2}),
\end{eqnarray}
or equivalently
\begin{eqnarray}\label{relation}
|\Omega_k|^3-\frac{6\pi^2M_\text{Pl}^2}{(n+\frac{1}{2})H_{g}^2}\Omega_\gamma|\Omega_k|-\frac{3\pi^2M_\text{Pl}^2}{2(n+\frac{1}{2})H_{g}^2}\Omega_m^2=0,
\end{eqnarray}
which is the quantum cosmological counterpart of the classical relation (\ref{yy}). { In ordinary quantum mechanics,  transition to excited states may occur induced through a ``time dependent'' term present in the Hamiltonian. But in General Relativity and subsequent quantum cosmology,  we have general covariance and general invariance of field equations. In our simple case, there is not any explicitly time dependent Lagrangian or Hamiltonian. Moreover, to have a change in the value of the quantum number, $n$, we would be need some  dynamics in the quantum cosmology to make such change. For example, in quantum mechanics, if we consider the superposition of states, then there is a possibility of time changing between various states on the superposition. But in strict quantum cosmology, we do not have any explicit time. Only through e.g., some quantum to classical transition and a decoherence process \cite{Kif}, whereby a WKB time may emerge, but only through the presence of fluctuations in the matter field, for example.   In our model,  there is an Hamiltonian constraint. In addition, because of the non-linearity of field equations (\ref{ss17}), the superposition of wave functions will  not be a solution of (\ref{ss17}). 

Let us further add that  in our model,  the quantum number $n$ is related to the matter content of universe, or equivalently, to the entropy of radiation
\cite{Sh1}. Hence,  changing the  value of $n$ would be equivalent to a change in the matter content of universe. But that in our model  is not consistent with covariant conservation of the fluid. But if  quantum matter fields would be instead present on a similar CdBB model we can investigate this aspect in an suitable context, which we leave for a subsequent work.} 

\section{CQHJ formulation for the FLRW  with a perfect fluid}

As we saw in Eq.(\ref{ss16}), the starting point of  CQHJ formulation is the insertion of the ansatz
\begin{eqnarray}\label{psi}
\Psi(q)=e^{iS(q)},
\end{eqnarray}
in the WDW equation (\ref{wdw}), where the wave function and the phase are analytically
extended to the complex plane by replacing  real coordinate of minisuperspace
$q$ with
 a complex coordinate
and keeping time (and lapse function)  real valued.
 By inserting  Eq.(\ref{psi})
into the WDW equation (\ref{wdw}) we obtain
\begin{eqnarray}\label{s1}
\frac{1}{2M}\left(\frac{dS}{dq}\right)^2+\frac{1}{2}M\omega^2q^2-i\frac{1}{2M}\frac{d^2S}{dq^2}-{\mathcal
E}=0.
\end{eqnarray}
The guidance
equation for complex quantum trajectories (\ref{ss19}) gives the momentum
\begin{eqnarray}\label{p}
p=\frac{M}{\tilde N}\frac{dq}{d\eta}:=\frac{dS}{dq}=-i\frac{d}{dq}\ln(\Psi),
\end{eqnarray}
where $(q,p)\in \mathbb{C}\times\mathbb C$.   As we saw in the previous section,
this means that the coordinate of
minisuperspace,  $q$, has been replaced by a complex variable $q=q_\text{R}+iq_\text{I}$,
where $q_\text{R}, q_\text{I}\in\mathbb{R}$.
 Using  relation (\ref{p}) in Eq.(\ref{s1}), we obtain the complex quantum super-Hamiltonian constraint
\begin{eqnarray}\label{s2}
\mathcal H_Q=\frac{p^2}{2M}+\frac{1}{2}M\omega^2q^2+Q(\Psi)-\mathcal E=0,
\end{eqnarray}
where 
\begin{eqnarray}\label{s3}
Q(\Psi)=\frac{1}{2Mi}\frac{dp}{dq}=-\frac{1}{2M}\frac{d^2}{dq^2}\ln(\Psi(q)),
\end{eqnarray}
denotes the complex quantum potential. Eq.(\ref{s2}) is a Riccati differential
equation for complex quantum momentum. The Hamilton equations of motion for the quantum state $\Psi_n$ can be derived from the quantum super-Hamiltonian (\ref{s2}) as (in gauge $\tilde N=1$)
\begin{eqnarray}\label{s4}
\begin{cases}
\dot q=\frac{\partial\mathcal H}{\partial p}=\frac{p}{M},\\
\dot p=-\frac{\partial\mathcal H}{\partial q}=-M\omega^2q-\frac{dQ}{dq},
\end{cases}
\end{eqnarray}
where a dot denotes derivative respect to $\eta$. Consequently, the complex
quantum
 Friedmann and the Raychaudhuri equations will be
\begin{eqnarray}\label{s5} 
\begin{cases}
 \frac{1}{2}M\dot q^2+\frac{1}{2}M\omega^2q^2+Q(\Psi_n)-\mathcal E=0,\\
 M\ddot q=-M\omega^2q-\frac{dQ(\Psi_n)}{dq}.
 \end{cases}
 \end{eqnarray}

\subsection{Trajectories in the CQHJ formulation}

Let us elaborate on how the CQHJ can be applied to extract solutions.

 { Before dealing with observable Universe, let us study in details two ground state and first excited universes.}  We start as a example with the ground state  universe,
 $n=0$, with eigenvalue ${\mathcal
 E}_0=\frac{\omega}{2}$, obtained from Eq.(\ref{eigen}).
 In this case, the  quantum potential (\ref{s3}) will be $Q=\frac{\omega}{2}$.
Moreover, from Eqs.(\ref{p}), (\ref{s4})  and $\Psi_0=C_0\exp(-\frac{M\omega}{2}q^2)$,
we obtain 
 \begin{eqnarray}\label{eq1}
 p=M\dot q=iM\omega q, 
 \end{eqnarray}
 with solution $q=q_\text{R}+iq_\text{I}=A\exp[i(\omega\eta+\theta)]$ where
 $A,\theta\in\mathbb{R}$.
 Note that the value of $A$, unlike the classical case, cannot be fixed by the quantum super-Hamiltonian
 (\ref{s2}). If we insert the quantum potential and complex conjugate variables $(q,p)$ into the constraint equation, it gives us only the eigenvalue of
ground state.  According to  Eq.(\ref{scale}), the real part of $q$ is related to the scale factor via $q_\text{R}=\frac{a}{a_g}-\frac{\Omega_m}{2|\Omega_k|}$.
Furthermore, the conformal time $\eta$ and cosmic time $t$ are related by $a_gH_gdt=a(\eta)d\eta$
as in the classical case, because of complex quantum variables are confined to
the minisuperspace, according to the quantization rule, and lapse function
and all time coordinates are real.  
Therefore, the scale factor will be
\begin{eqnarray}\label{f1}
\begin{cases}
a(t)=a_g\left(A\cos(\omega\eta+\theta)+\frac{\Omega_m}{2|\Omega_k|}\right),\\
t=\frac{1}{H_g}\left(\frac{A}{\omega}\sin(\omega\eta+\theta)+\frac{\Omega_m}{2|\Omega_k|}\eta\right).
\end{cases}
\end{eqnarray}
From (\ref{f1}), using initial conditions $a(t_g)=a_g$ and $H_g=H(t_g)=\frac{1}{a}\frac{da}{dt}|_{t=t_g}$,
we obtain
\begin{eqnarray}\label{A}
A=\frac{1}{\sqrt{|\Omega_k|}}\sqrt{1-\Omega_k-\Omega_m-\Omega_\gamma+\frac{H_g^2\Omega_k^2}{12\pi^2M_\text{Pl}^2}}.
\end{eqnarray}
Note that according to the Eq.(\ref{relation}), $1-\Omega_k-\Omega_m-\Omega_\gamma\neq0 $ in quantum cosmology. We also can find the following relation between the scale factor and the
imaginary part as
\begin{eqnarray}\label{ppp}
\left(\frac{a}{a_g}-\frac{\Omega_m}{2|\Omega_k|}\right)^2+q_\text{I}^2=A^2.
\end{eqnarray}
 A  point to be noticed is that the real part of the scale
factor obtained in Eq.(\ref{f1}) is similar to the classical motion of the 
closed Universe (\ref{cl1}),  but in the quantum derived expression for the universe,  (\ref{f1}), the maximum of
scale factor is given by $a_\text{Max}:=a_{g}A+\frac{a_g\Omega_m}{2|\Omega_k|}$
and the imaginary part of motion is not negligible at all and a universe with
$n=0$ is entirely in quantum domain.
Another interesting feature is that
\begin{eqnarray}\label{q0}
\frac{1}{2}M\omega^2q^2+\frac{p^2}{2M}=0, \,\,\,Q=\frac{\omega}{2},
\end{eqnarray}
which indicates that the dynamics of such universe is completely originated
from the quantum potential. The solution $q=q_\text{R}+iq_\text{I}=A\exp[i(\omega\eta+\theta)]$
 shows that the the zero-mode universe is not singular. Moreover, we can easily show that the real part of the solution (\ref{f1})
is not singular for $\Omega_m>1$. 
 We will show that for universes with very large values of quantum
number  $n$, the quantum potential vanishes and the model reduce to a classical
Universe. 

Before dealing with this classical limit of our model, let us consider a universe with $n=1$ as a second example. 

To obtain the trajectory for the $n=1$ quantum universe,
we apply $\Psi_1=C_1y\exp(-\frac{M\omega}{2}q^2)$ to the definition of momentum
in Eq.(\ref{p}) which leads 
\begin{eqnarray}\label{q20}
p=M\dot q=i(M\omega q-\frac{1}{q}).
\end{eqnarray}
The integration gives the eigen-trajectory
\begin{eqnarray}\label{qq2}
M\omega q^2=1+Ae^{i(2\omega\eta+\theta)},
\end{eqnarray}
where $A,\theta\in\mathbb{R}$.
The real part of (\ref{qq2}) together the relation  $q_\text{R}=\frac{a}{a_g}-\frac{\Omega_m}{2|\Omega_k|}$ gives 
\begin{eqnarray}\label{q3}
\begin{array}{cc}
\frac{a}{a_g}=\frac{\Omega_m}{2|\Omega_k|}+\\\frac{1}{\sqrt{2M\omega}}
\left(1+A\cos(2\omega\eta+\theta)+\sqrt{A^2+1+2A\cos(2\omega\eta+\theta)}\right)^\frac{1}{2}.
\end{array}
\end{eqnarray}
The value of $A$ can be calculated from initial values $a(t_g)=a_g$ and
$H(t_g)=H_g$ similarly to the case of ground state.

A more complete understanding of the dynamics in complex minisuperspace is gained from the consideration of the complex  Raychaudhuri equation. Inserting $\Psi_1$
into  Raychaudhuri equation in (\ref{s5}) gives
\begin{eqnarray}\label{psi1}
M\ddot q=-M\omega^2q+\frac{1}{M}q^{-3}.
\end{eqnarray}
When $|q|\gg\frac{1}{\sqrt{M\omega}}$,
the quantum force $-\frac{dQ}{dq}=\frac{1}{Mq^3}$ approaches to zero and the classical equation of motion is recovered. When $|q|\ll\frac{1}{\sqrt{M\omega}}$, the classical force becomes negligible and the motion is dominated by the quantum force.   Fig.(1) shows the complex paths in minisuperspace
for $n=1$ universe. This universe is non-singular like as the $n=0$ universe.
\begin{figure}[htb]\label{fig}
\begin{center}
\includegraphics[width=6cm]{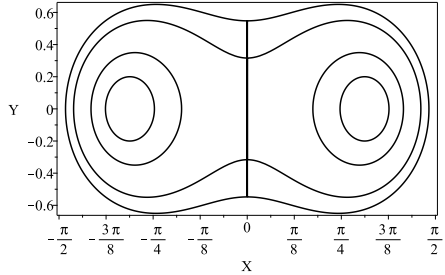}
%\hfill
%\centering
%\includegraphics[width=10cm]{1.eps}
\caption{Complex paths in complex minisuperspace for $n=1$ where we defined $X:=\sqrt{M\omega}q_\text{R}$
and $Y:=\sqrt{M\omega}q_\text{I}$. The contours are plotted for $\omega=0.5$, $A=0.4, 0.7,
1.1$ and $1.3$ values.  }
\end{center}
\end{figure}

\subsection{Emergence of a classical Universe}

Let us now investigate the behaviour of the  model for very large values of the quantum state
number $n$. Let us first estimate the value of quantum number $n$ for our
Universe.

In Ref.  \cite{Sh1}, it was  showed that the total entropy for  radiation in the  model
investigated herein is given by \begin{eqnarray}\label{entropy}
S_\gamma=1.3g^\frac{1}{4}\left(n+\frac{1}{2}\right)^\frac{3}{4},
\end{eqnarray} 
where $g$ is the internal degrees of freedom of radiation. This shows that
the value of quantum number $n$ is related to the  matter content
of Universe. For  created universe from nothing with large amount of matter, the quantum
number is also large, and inversely, for a universe with large quantum number
$n$, the matter content of that universe is also great. 
On the other hand, the entropy of radiation in the observable part
of Universe is about $10^{88}$ \cite{Entropy}. Consequently, this allow us
to estimate the value of the quantum  number  in our Universe as $n\gtrsim\ 10^{118}$.

For  a large quantum number $n$, the wave function (\ref{eigen}) has the following
asymptotic expansion
\begin{eqnarray}\label{H1}
\Psi_{2n+1}(y)=C_{2n+1}\sin\left(\sqrt{2Mn\omega}y\right).
\end{eqnarray}
Therefore, the complex quantum potential (\ref{s3}) will be
\begin{eqnarray}\label{q2}
Q=\frac{2n\omega}{\sin^2(\sqrt{2Mn\omega}q)}.
\end{eqnarray}
Moreover, using  (\ref{p}) and (\ref{s4}), the  equation of motion
will be
\begin{eqnarray}\label{b1}
M\dot q= \begin{cases}
i\sqrt{2Mn\omega}\cot\left(\sqrt{2Mn\omega}q\right),\sqrt{2Mn\omega}|q|\gg1,\\
i\left(2Mn\omega q-\frac{1}{q}\right),\hspace{1.4cm}\sqrt{2Mn\omega}|q|\ll1.
\end{cases}
\end{eqnarray}
The integration for  large values of $q$ gives the eigen-trajectory as
\begin{eqnarray}\label{m1}
\cos\left(\sqrt{Mn\omega}q\right)=Ae^{-i(2n\omega\eta+\theta)},
\end{eqnarray}
where $A,\theta\in\mathbb{R}$.
Separating the real and imaginary parts of $q=q_{\text R}+iq_{\text I}$ gives us
\begin{eqnarray}\label{b2}
\begin{cases}
e^\beta\cos\alpha =2A\cos(2n\omega\eta+\theta),\\
e^\beta\sin\alpha =2A\sin(2n\omega\eta+\theta),
\end{cases}
\end{eqnarray}
where $\alpha=\sqrt{2Mn\omega}q_{\text R}$ and $\beta=\sqrt{2Mn\omega}q_{\text{I}}$.

The solution of the above equations, using the definition of $M$ and $\omega$ in
(\ref{pp}) and the relation between conformal time and comoving time, $a_gH_gdt=ad\eta$, yields 
\begin{eqnarray}\label{ff1}
\frac{a(t)}{a_{g}}=\left(\frac{2n}{3}\right)^\frac{1}{4}\left(\frac{H_g^2|\Omega_k|}{\pi M_{\text{Pl}}}\right)^\frac{1}{2}\sqrt{t},
\end{eqnarray}
and
\begin{eqnarray}\label{ff2}
q_{\text I}=\frac{\ln(2A)}{\sqrt{4nM\omega}}.
\end{eqnarray}
Equation (\ref{ff2}) implies that for very large values of quantum number, $n\gg1$,  the imaginary part of the motion vanishes and the trajectory falls entirely on the real axis of minisuperspace. Inserting  the initial condition $H(t_g)=H_g$ in the time derivative of Eq.(\ref{ff1}) gives $2H_gt_g=1$. Furthermore, the  initial condition $a(t_g)=a_g$ gives
\begin{eqnarray}\label{fa1}
|\Omega_k|=\frac{\pi M_\text{Pl}}{H_g}\left(\frac{6}{n}\right)^\frac{1}{2}=2\pi
\frac{t_g}{t_\text{Pl}}\left(\frac{6}{n}\right)^\frac{1}{2},
\end{eqnarray}
where $t_\text{Pl}=1/M_\text{Pl}$ denotes Planck's time in natural units. If we insert
Eq.(\ref{fa1}) into (\ref{relation}) we will obtain the energy density parameter
%of monopoles as
\begin{eqnarray}\label{fa2}
\Omega_m=\left(\frac{8\pi t_g}{t_\text{Pl}}(1-\Omega_\gamma)\right)^\frac{1}{2}\left(\frac{6}{n}\right)^\frac{1}{4}.
\end{eqnarray}
 On the other hand, for $\sqrt{2Mn\omega}|q|\ll1$ the motion is dominated by the quantum force at the very early Universe, where according to the first
equation in Eq.(\ref{b1}), the eigen-trajectory for very small values of $q$, similar to the (\ref{q3})  is oscillatory and non-singular in an initial moment, $t=0$, with minimum 
\begin{equation}\label{min-eq}
a(0)=\frac{\Omega_m}{2|\Omega_k|}a_g.
\end{equation}
Inserting the above  obtained $|\Omega_k|$ and $\Omega_m$ in  Eqs.(\ref{fa1}) and (\ref{fa2}) gives
\begin{eqnarray}\label{fa3}
 \Omega_\gamma=1-\frac{2\pi t_g}{t_\text{Pl}}\left(\frac{a(0)}{a_g}\right)^2\left(\frac{6}{n}\right)^\frac{1}{2}.
 \end{eqnarray}
 Inserting again the energy density parameter of radiation obtained in above
 equation into Eq.(\ref{fa2}) gives
 \begin{eqnarray}\label{fa4}
 \Omega_m=\frac{4\pi t_g}{t_\text{Pl}}\frac{a(0)}{a_g}\left(\frac{6}{n}\right)^\frac{1}{2}.
 \end{eqnarray}

\subsection{Classical implications from CQHJ}

The grand unification epoch could have ended at approximately $t_g\simeq10^{-36}$ seconds after the Big Bang. Moreover, the quantum  description of the Universe is that of a non-singular scenario with initial scale factor $a(0)$ as indicated in (\ref{min-eq}).

\subsubsection{On the flatness issue}

If we assume the initial value of scale factor at the beginning of Planck's
length, $a(0)\simeq10^{-33}$ cm, and take $n\gtrsim10^{118}$ as estimated in previous
section, then, according to Eq.(\ref{fa1}) the  spacial curvature  parameter  at the  GUT phase transition time
will be  
\begin{eqnarray}\label{la1}
|\Omega_k|\lesssim10^{-58}.
\end{eqnarray}
Using the definition of curvature parameter $|\Omega_k|=\frac{1}{a_g^2H_g^2}$,
relation $2H_gt_g=1$ and Eq.(\ref{la1}) we obtain the linear size of Universe
at GUT time
\begin{eqnarray}\label{la2}
a_g\simeq1\,\,\text{mm}.
\end{eqnarray}
Now, inserting these values into Eqs.(\ref{fa3}) and (\ref{fa4}) give us
\begin{eqnarray}\label{la3}
\Omega_m\lesssim10^{-78},\hspace{.2cm}\Omega_\gamma\simeq1-10^{-86}.
\end{eqnarray} 
In other words, according to  Eq.(\ref{b1}), for very large values of $n$
and $\sqrt{2M\omega n}|q|\ll1$,
or equivalently, for  scale factors smaller than
$a_g$, the model predicts an oscillating quantum Universe, where the minisuperspace is complex and
 without initial Big Bang singularity, while for $a\gtrsim\ a_g$ the emerged
Universe
is completely classical with real minisuperspace, very close to
spatially flatness, radiation dominated with scale
factor
given by\begin{eqnarray}\label{scale1}
a(t)=a_g\sqrt{\frac{t}{t_g}},
\end{eqnarray}
 where the density of 
matter
%monopoles are
is 
lower by many orders of magnitude.

\subsubsection{On the horizon issue}

 According to the CMB observations the whole of Universe was causally connected at last scattering surface time \cite{Mukhanov}. But in standard FLRW classical  cosmology, the Universe is causally connected by an angle of order unity,  which is in conflict with observation.

The necessary condition for the universe to be causally connected at time $t$ is
\begin{eqnarray} \label{horizon}
%\begin{array}{cc}
d_{H}(t)=a(t)\int_{0}^{t} \frac{dt'}{a(t')}\geq d_{p}(t)=a(t)\int_{0}^{r_\text{Max}}\frac{dr}{\sqrt{1-kr^2}},
%\end{array}
\end{eqnarray}
where $d_{H}$ and $d_{p}$ represent the horizon and proper distances at time $t$ respectively.

For open and flat universes the right hand side of Eq.(\ref{horizon}) becomes infinite for $r_\text{Max}=\infty$ and therefore the globally causality failed at any finite time. On the other hand, for a closed universe, $r_\text{Max}$ is finite and we can define a causal time $(t_\text{cau})$ at which the whole of Universe becomes causally connected as
\begin{eqnarray} \label{time}
\int_{0}^{t_\text{cau}} \frac{dt'}{a(t')}=\int_{0}^{1}\frac{dr}{\sqrt{1-r^2}}=\frac{\pi}{2}.
\end{eqnarray} 
To calculate the left-hand side of (\ref{time}) we assume that the classical
Universe started just after the moment $t_g$, and consequently take the lower bound
of integration as GUT time. Hence, we will obtain\begin{eqnarray} \label{time3}
t_\text{cau}=(\frac{\pi}{4}+1)^{2} t_{g},
\end{eqnarray}
which shows  that the whole of Universe becomes causally connected at GUT phase
transition time, because of quantum effects of gravity at very early Universe, in this rather specific and simple model we have been exploring.

\section{Conclusions and Discussion}

In this paper, we have introduced a simple quantum cosmological model to
which applied the quantum
Hamilton-Jacobi formalism with the concept of a complex quantum trajectory \cite{Complex}.

Our purpose was to address from a new and different perspective some problems of the standard  Big Bang setting of cosmology. This scenario based on matter
content being described by dust and radiation is observationally successful in describing
the present epoch of Universe and up to sometime into the past. From the microwave background
radiation it is possible to trace it up to a red-shift $z\sim10^3$,
while nucleosynthesis probes it up to $z\sim10^{11}$. We do not have observational
evidence regarding the correctness of this scenario at larger red-shifts, for example
the standard GUT area, $z\sim10^{27}$. 

On the other hand,
 theoretical  inconsistencies of scenario, like the existence of an initial
singularity and also the flatness and horizon problems, definitely
suggests the breakdown of this framework at some large red-shifts. Cosmological inflation
is a mechanism that  improves on those mentioned problems.

In this paper, we have nevertheless explored on the ability of quantum cosmology to provide a new insight on those  problems, without
inserting explicitly any new set of fields,  parameters or  extensions. Our
new tool is the CdBB framework.

We considered a closed FLRW Universe filled
with a dust fluid and radiation. We showed that for very large values of a quantum state number $n$, which
according to Eq.(\ref{entropy}) is related to the entropy of radiation in the
Universe, the classical
Universe can emerge from an  oscillating complex quantum Universe, without singularity, horizon  and flatness problems. 

{ On a final note, let us add that it would be of significant interest to extend this work to other models. Namely, either anisotropic\footnote{Viz. Bianchi type-I or even type-IX, to discuss eventual emergent chaotic behaviour \cite{C}.} or with a scalar field. Furthermore, within a setting in which inhomogeneities are allowed to be present perturbatively. Dealing with inhomogeneous perturbations will be of relevance because different interpretations of quantum mechanics may have different  observational consequences. Specifically, if choosing to employ (in the central role)  the ``collapse of the wave function" towards  the prediction of the spectrum of perturbations (cf. in particular \cite{A}). As far as the usual  dBB approach to quantum cosmology is concerned  linear cosmological perturbations have been considered (see
references in \cite{Fab}). Falsifiable observational consequences were pointed and some fitted with known data, although others remain to be tested. In
dBB theory of quantum cosmology, the desirable fluctuations (inhomogeneities
in the matter fields densities) do occur, and the undesirable fluctuations
(Boltzmann brains in the late Universe) presumably do not occur, because
there are no external observer causing the wave function to collapse \cite{G}.  Regarding CdBB into quantum cosmology, as introduced and  explored in this paper, that remains an open issue to contemplate. 
We are leaving the above enticing research lines for future works. }
\\

\section{Acknowledgments} The authors would like to sincerely thank the
anonymous referees for constructive and helpful comments to improve
the original manuscript.\\

P.V. Moniz research work is supported by the Portuguese grant UID/MAT/00212/2013.

\end{document}